\documentclass[12pt]{article}
\usepackage{amssymb}
\usepackage{graphicx}
\usepackage{graphics}
\usepackage{amsmath}
\usepackage{url}

\def\e{{\rm e}}

\def\l{\left(}
\def\r{\right)}

\newcommand{\be}{\begin{equation}}
\newcommand{\ee}{\end{equation}}
\newcommand{\ba}{\begin{align}}
\newcommand{\ea}{\end{align}}
\newcommand{\bg}{\begin{gather}}
\newcommand{\eg}{\end{gather}}
\newcommand{\bseq}{\begin{subequations}}
\newcommand{\eseq}{\end{subequations}}

\begin{document}

\title{Constraining neutrino superluminality from searches for
  sterile neutrino decays}

\author{
D.\,S.\;Gorbunov\thanks{{\bf e-mail}: gorby@ms2.inr.ac.ru},
E.\,Ya.\;Nugaev\thanks{{\bf e-mail}: emin@ms2.inr.ac.ru}
\\
{\small{\em
Institute for Nuclear Research of the Russian Academy of Sciences,
}}\\
{\small{\em
60th October Anniversary prospect 7a, Moscow 117312, Russia
}}\\
}
\date{}

\maketitle

\begin{abstract}
Superluminal neutrinos are expected to lose energy due to
bremsstrahlung. It is dominated by $e^+e^-$-pair production if kinematically
allowed. The same signature was used in searches for 3-body decays of
hypothetical heavy sterile neutrinos. From the 
published analyses of these searches performed by 
CERN PS191 and CHARM experiments we set upper limits on the
neutrino velocity in the energy range from 0.2\,GeV to 280\,GeV. Our
limits are well below the neutrino
velocity favored by the recent OPERA results. For energy-independent
neutrino velocity the limits obtained in this paper are stronger than those coming
from ICARUS experiment and observations of Supernova SN1987a.   
\end{abstract}


Recently neutrino physics has attracted renewed attention because
of the surprising result of the OPERA
collaboration\;\cite{Adam:2011zb} (see, however\;\cite{Sirri}).
Their claim is that the travel time of muon neutrino $\nu_\mu$ with average
energy  $17.5$ GeV is smaller than the light travel time at
the confidence level above 5\,$\sigma$.  This result was interpreted as
a superluminal propagation of muon neutrino, which can be parametrized 
by 
\begin{equation}
\delta \equiv\l {v_\nu}^2-1\r\simeq 5\times 10^{-5}\;,
\label{OPERA}
\end{equation}
where $v_\nu$ is the neutrino velocity and we set the Planck constant $\hbar$
and photon velocity $c$ equal to unity, $\hbar=c=1$.

However,  Cohen and Glashow \cite{Cohen:2011hx} put forward a serious objection to this superluminal
interpretation. They  considered several decay processes
which would take place in vacuum under the assumption $\delta>0$. It was pointed
out that by analogy to the Cherenkov radiation, the electroweak process
\begin{equation}
\nu_{\mu}\to\nu_{\mu}+e^{+}+e^-
\label{nutonuee}
\end{equation}
should occur, and the neutrino spectrum should be drastically
changed as neutrinos travel  a distance of
 about 730 km from the target at CERN to the OPERA
detector at Gran Sasso. 
The neutrino energy spectrum near the OPERA detector was
measured by the ICARUS collaboration, and no attenuation was observed.  
Using the result of Cohen and Glashow
\cite{Cohen:2011hx} and the analysis of the ICARUS data, the limit
\begin{equation}
\delta<4\times 10^{-8}
\label{ICARUS}
\end{equation}
was obtained \cite{ICARUS:2011aa} under the assumption of
energy-independent $\delta$. 



In this paper we give interpretation of other neutrino experiments,
where the emission of $e^+e^-$-pair could be observed,
in terms of the decay~(\ref{nutonuee}).
The rate of the decay (\ref{nutonuee}) in the laboratory frame is given by
\begin{equation}
\Gamma=k'\frac{G_F^2}{192\,\pi^3}E^5\delta^3\;,
\label{rate}
\end{equation}
where $G_F$ is the Fermi constant, 
$E$ is the neutrino energy and $k'$ is a numerical constant. In the
original paper \cite{Cohen:2011hx} the value $k'=1/14$ was used for
estimates.  As shown in Ref. \cite{Bezrukov:2011qn}, this value is
model-dependent and can be made two or three times smaller.
Due to the very
strong dependence of (\ref{rate}) on $\delta$ and neutrino energy
the precise value of $k'$ is not crucial for our estimates and we
adopt $k'=1/14$, which was also accepted in Ref. \cite{ICARUS:2011aa} when
obtaining the upper bound (\ref{ICARUS}).  
The threshold of the pair production 
(\ref{nutonuee}) is 
\begin{equation}
\label{threshold}
E=2m_e/\sqrt\delta\;.
\end{equation}

The signature of the process (\ref{nutonuee}) is the appearance of
$e^+e^-$-pair from nothing downstream of a neutrino beam. Remarkably,
the same signature is exploited in searches for heavy sterile neutrino
decays. Indeed, sterile neutrinos 
mix with the Standard Model (SM)
active neutrinos thereby producing masses and mixing in active
neutrino sector required to explain the neutrino oscillations. This
mixing also gives rise to sterile neutrino production in meson
decays and subsequent sterile neutrino decays into SM particles 
including
$\e^+e^-\nu$, i.e. the same final state as in (\ref{nutonuee})
 (for a
concrete model example see, e.g., Ref. \cite{Gorbunov:2007ak}).

\begin{figure}[!htb]
\hskip 0.1\textwidth
\includegraphics[width=0.8\textwidth]{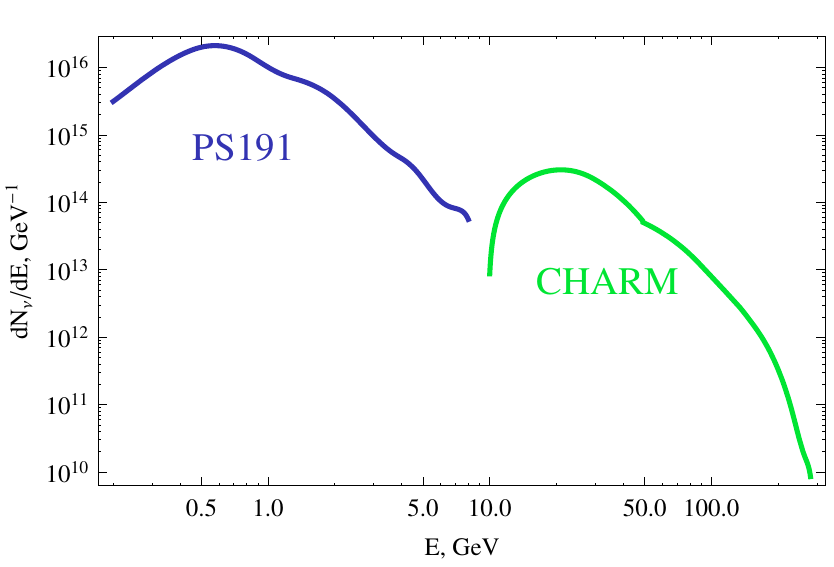}
\caption{Neutrino spectra in CERN PS191 and CHARM experiments. 
\label{Fig-spectra}
}
\end{figure}

In the neutrino energy range $E\sim$\,1-8\,GeV 
the stringent bound on the
lepton pair production 
follows from the analysis of CERN PS191
experiment. Neutrinos were eventually produced on a beryllium
target by protons of energy 19\,GeV. 
This experiment was specially designed
\cite{Bernardi:1985ny} to look for possible sterile neutrino decays to
$e^+e^-\nu$ in vacuum and has 
observed 
no events with collected statistics of
$8.6\times 10^{18}$ protons on target. 
The differential energy spectrum of
muon neutrinos $dN_\nu/dE$ expected from Monte Carlo simulations
in the range of energies from $200$\,MeV to $8$\,GeV is
presented in Ref. \cite{Bernardi:1986hs} and shown in
Fig.\,\ref{Fig-spectra}.  
In the case of process (\ref{nutonuee}),
neutrino loses about three-quarters of its energy in
each emission \cite{Cohen:2011hx}, so even for  $E\sim 200$ MeV the
energy of leptons is well above the ionization threshold of the
detector, and the efficiency of registration was close to  $100\%$  \cite{Bernardi:1986hs}.

\begin{figure}[!htb]
\hskip 0.1\textwidth
\includegraphics[width=0.8\textwidth]{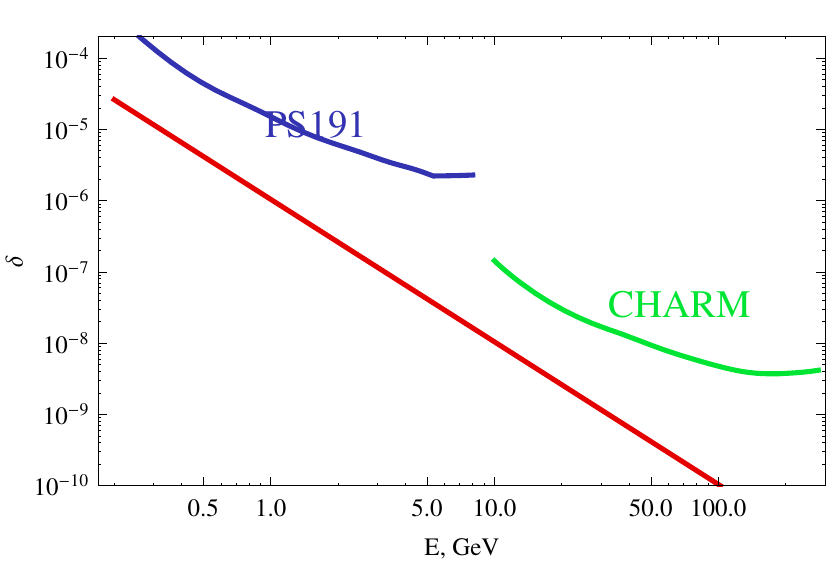}
\caption{Upper limits on $\delta\l E\r$ from the results 
of CERN PS191 and CHARM experiments; 
straight line refers to the pair-production threshold
\eqref{threshold}, 
so the region below this line is kinematically forbidden.     
\label{Fig-limits}
}
\end{figure}

For the vacuum camera of length $l\approx10$\,m each neutrino has a
probability of $e^+e^-$-pair production
\begin{equation}
\label{probability}
P=P\l E\r = l\cdot \Gamma\approx 0.8\times 10^{-13}\times \l
\frac{\delta}{10^{-5}}\r^3 \l \frac{E}{1\,{\rm GeV}}\r^5\;. 
\end{equation}
For a given neutrino energy interval one writes for the number of
$e^+e^-$-pairs from nothing: 
\begin{equation}
\label{number-of-events}
N=\int \!\!dE\; \frac{dN_\nu}{dE}\,P\!\l E\r\;.
\end{equation}
In the original search for sterile neutrino\,\cite{Bernardi:1985ny} 
the obtained limit was based on a set of requirements on the signal 
events, one of which is the absence of any double track events due to the
$e^+e^-$ pair. In our case the angle between electron and positron 
is small, $\sim\sqrt{\delta}$, well below the detector angular
resolution, $\sim 10^{-2}$, see Fig.\,\ref{Fig-limits}. 
To constrain neutrino superluminality we use the subset DST2 of $\sim
4000$ events 
which (judging on description presented in Ref.\,\cite{Levy}) includes
our signal events (if any): one expects a single track with double heat in
electromagnetic calorimeter. Having no real data in hand, we cannot
proceed further in analysis of DST2. Then, to obtain the limit on
$\delta$, we  
do not subtract SM background and suppose
that all these $\sim 4000$ are due to the Cohen-Glashow  effect. One can
set an upper limit on $\delta$ at
95\% C.L. by adopting the normal distribution and requiring $N\lesssim4100$.
Then for the energy-independent $\delta$ we integrate in
r.h.s. of Eq.\,\eqref{number-of-events} over the entire interval of
neutrino energies 0.2--8\,GeV and obtain
\[
\delta<2.2\times10^{-6} ~~{\rm at}~95\%\,{\rm CL}\;.
\] 

In the case of energy-dependent $\delta=\delta\l E\r$ one has to integrate
in r.h.s. of Eq.\,\eqref{number-of-events} to obtain the limits on 
neutrino dispersion relation. In particular, if $\delta\l E\r$ does
not vary significantly in the energy interval $E\pm E/2$, one
can exploit the relation 
\[
4100\gtrsim N_E\equiv \int_{E-E/2}^{E+E/2} \!\!dE'\;\frac{dN_\nu}{dE'}\,P\!\l
E'\r \approx l \delta^3\l E\r\,k'\frac{G_F^2}{192\,\pi^3} 
\int_{E-E/2}^{E+E/2} \!\!dE'\;\frac{dN_\nu}{dE'}\,E^{'5}
\]
to set a limit on moderately-varying $\delta\l E\r$ as presented in 
Fig.\,\ref{Fig-limits}.  

We obtain similar limits from the negative results of CHARM experiment
\cite{Bergsma:1983rt} which also performed
searches for  sterile neutrino decays.  
In this experiment, 
the CHARM detector of length $l\approx 12$ m was
exposed to wide-band neutrino beam produced by $\sim 1.4\times 10^{18}$ protons with 
energies $\sim 400$\,GeV 
incident 
on a copper target.
The neutrino spectrum expected from Monte Carlo simulations\footnote{We use 
wide-band neutrino beam spectrum presented in \cite{Nobel-Jack} for CDHS
detector, which was placed right in front of the CHARM detector along the neutrino beam. The effective areas of these detectors were almost identical. We thank A. Rozanov for this suggestion.
} 
is given in Ref. \cite{Nobel-Jack} and shown in Fig.\,\ref{Fig-spectra}. 
The selected events were required to have a shower energy $W$ deposited in the CHARM
calorimeter between $7.5$ and $50$ GeV and a value of the variable $W^2\theta^2$ below $0.54$ GeV$^2$ ($\theta$ is the angle between the shower axis and the direction of incoming neutrino). A total of $331$ events were selected. The 
number of events attributed to heavy neutrino decay is compatible with zero, $1\pm 41$ event. In the case of decay (\ref{nutonuee}) one has $\theta^2\sim\delta$,  
so the selected above cuts do not interfere in our analysis. 
To place a limit on $\delta$ in this case we require to have less than $85$ events
due to the bremsstrahlung (\ref{nutonuee}).
Also, the statistics for $e^+e^-$ production from nothing was approximately doubled 
when detector was exposed to muon antineutrino flux produced by $\sim 5.7\times 10^{18}$ protons on target \cite{Bergsma:1983rt}.
For energy-independent 
$\delta$ we obtain from \eqref{number-of-events} after integrating
over the whole available neutrino energy range 10--280\,GeV, and taking
into account the $44\%$ efficiency \cite{Bergsma:1983rt} of $e^+e^-$-registration, 
\[
\delta<3.6\times10^{-9}.
\] 
This limit is stronger than the constraint \eqref{ICARUS} obtained by
the ICARUS collaboration. 
Strikingly, this  limit is even comparable to the
stringent direct bound \cite{Longo:1987ub} obtained 
from the observations \cite{Hirata:1987hu} 
of neutrino signal from Supernova 1987a, 
\begin{equation*}
\delta<4\cdot 10^{-9}\;, 
\end{equation*}
which is valid for the energy range around 10\,MeV. However, our limit
is weaker than what 
is expected
\,\cite{Cohen:2011hx} 
from the analysis of 10-100\,TeV 
neutrino events measured by the IceCube experiment. The reason is the
strong dependence of the emission rate\,\eqref{rate} on neutrino energy.

For a moderately-varying $\delta\l E\r$ we follow the procedure
used for PS191 and obtain limits presented in Fig.\,\ref{Fig-limits}. 



To summarize, in this paper we have interpreted the negative results of
searches for sterile neutrino decays in CERN PS191 and CHARM
experiments as the restriction on $e^+e^-$-pair production by possibly
superluminal muon neutrinos. We have set the upper limits on
neutrino velocity in the energy ranges 0.2--8\,GeV and 10--280\,GeV 
supposing neutrino velocity to be only moderately varying with energy.
In a particular model the limit can be refined by convolution of neutrino
production rate with neutrino beam spectra according to formulas presented in the paper.
Covering the gap
between 8 and 10\,GeV should be possible with information on high energy end of
the neutrino spectrum
in CERN PS191 experiment. 
In both experiments further strengthening 
of the bound might be possible 
with 
adjustment 
of the selected cuts.
In particular with lower cut on $W^2\theta^2$ in CHARM experiment.
Potentially interesting limits one can expect from reanalysis of data collected
at $35$-meter length detector with CHARM calorimeter module installed, which was designed to search for heavy neutrinos produced by charmed meson decays \cite{Bergsma:1983rt}.

Analogous limits are expected to be obtained by the dedicated analysis
of the data from NOMAD experiment, see Ref.~\cite{Cattaneo:2011wg} for the preliminary bound.

We thank V.\,Rubakov and S.\,Sibiryakov for discussions,   
and A.\,Rozanov and F.\,Vannucci for correspondence.

The work is supported in part by the grant of
the President of the Russian Federation NS-5590.2012.2, by Russian
Foundation for Basic Research grants 11-02-01528-a (D.G.) and
11-02-92108-YAF\_a (D.G.) and by the SCOPES grant (D.G.). 


\end{document}